# Noise constrained least mean absolute third algorithm

Sihai GUAN [1], Zhi LI [1]

**Abstract:** The learning speed of an adaptive algorithm can be improved by properly constraining the cost function of the adaptive algorithm. Besides, the stabilization of the NCLMF algorithm is more complicated, whose stability depends solely on the input power of the adaptive filter and the NCLMF algorithm with unbounded repressors is not mean square stability even for a small value of the step-size. So, in this paper, a noise variance constrained least mean absolute third (LMAT) algorithm is investigated. The noise constrained LMAT (NCLMAT) algorithm is obtained by constraining the cost function of the standard LMAT algorithm to the third-order moment of the additive noise. And it can eliminate a variety of non-Gaussian distribution of noise, such as Rayleigh noise, Binary noise and so on. The NCLMAT algorithm is a type of variable step-size LMAT algorithm where the step-size rule arises naturally from the constraints. The main aim of this work is first time to derive the NCLMAT adaptive algorithm, analyze its convergence behavior, mean square error (MSE), mean-square deviation (MSD) and assess its performance in different noise environments. Finally, the experimental results in system identification applications presented here illustrate the principle and efficiency of the NCLMAT algorithm.

## 1. Introduction

Adaptive filter (AF) algorithm is frequently employed in equalization, active noise control, acoustic echo cancellation, biomedical engineering and many other fields [1][2]. In many communication applications, additive white Gaussian noise is an appropriate model for thermal and possibly interference noise, and the noise power can be accurately estimated prior to channel tap estimation. Hence, the investigation of the benefits of incorporating noise variance knowledge into an adaptive channel estimation algorithm is of some practical interest in recent decades, noise as a constraint has been studied, for example, the noise constrained LMS (NCLMS) algorithms [3][4]. Besides, although the NCLMS-type algorithms have many advantages, we may have worse choice for system identification with noise constraint in some case where the measurement noise is non-Gaussian [5]. So, the noise constrained LMF (NCLMF) algorithm was proposed [6]. But, the stabilization of the LMF algorithm is more complicated than that of the LMS algorithm, whose stability, for a given step-size, depends solely on the input power of the adaptive filter and the LMF algorithm with unbounded repressors is not mean square stability even for a small value of the step-size [7]. Significantly, the LMAT algorithm is built on the minimization of the mean of the absolute error value to the third power. The error function is a perfect convex function with respect to filter coefficients, so there is no local minimum for the LMAT algorithm [8].

✉ Zhi Li
15107739422@163.com

Sihai Guan
gcihey@sina.cn

[1] School of Electro-Mechanical Engineering, Xidian University, N0. 2 South Taibai Road, Xi’an, 710071, China

Using a constraint would make the step size in the LMAT algorithm act as a variable step size. For the research of variable step-size LMAT algorithm, Guan S and Li Z proposed an NVSLMAT algorithm [9]. In the NVSLMAT algorithm, fewer parameters need to be set. But this algorithm not used the additive noise variance. Ultimately, on the basis of work in [6][8], we propose a noise constrained LMAT (NCLMAT) algorithm by constraining the cost function of the standard LMAT algorithm to the third-order moment of the additive noise variance. Besides, the mean convergence, steady state MSE and MSD of the NCLMAT algorithm are derived. The computational complexity of the NCLMAT algorithm is analyzed theoretically. Besides, its characteristics are evaluated in different noise environments. In order to further illustrate that the NCLMAT algorithm is superior to the LMAT and NCLMF algorithms, a number of simulation results are carried out to corroborate the theoretical findings, and as expected, improved performance is obtained through the use of this technique.

The paper is organized as follows: the proposed NCLMAT algorithm is introduced in Section 2. The performance of the NCLMAT algorithm is explored in Section 3. The numerical simulations are carried out in Section 4 and the conclusion is drawn in Section 5.

## 2. The proposed NCLMAT algorithm

The output of an FIR channel in the presence of an additive noise can be written as

$$d_k = \mathbf{w}_{opt}^{\mathrm{T}} \mathbf{x}_k + \xi_k \tag{1}$$

where $\mathbf{w}_{opt} = [w_0, w_1, \cdots, w_{N-1}]^{\mathrm{T}}$ corresponds to a channel/system impulse response with $N$ taps. $N$ is the filter length. The input data vector of the unknown system at time instant $k$ and the observed output signal are denoted by $\mathbf{x}_k = [x(k), x(k+1), \cdots, x(k+N-1)]^{\mathrm{T}}$ and $d_k$, respectively. $\xi_k$ is a zero-mean i.i.d. process with an arbitrary probability density function. It has a variance of $\sigma_\xi^2$. We consider real-valued quantities for simplicity.

So

$$e_k = d_k - \mathbf{w}_k^{\mathrm{T}} \mathbf{x}_k \tag{2}$$

The cost function to be minimized is given by

$$J(\mathbf{w}) = \mathrm{E}\left[e_k^m\right], \ for \ m > 0 \tag{3}$$

Minimization of the cost function (3) over $\mathbf{w}_k$ gives the optimal weight value at time $k$, i.e., $\mathbf{w}_k = \mathbf{w}_{opt}$ with $J(\mathbf{w}_{opt}) = J_{\min}$, where $J_{\min}$ is the value of the cost function (3) evaluated at $\mathbf{w}_k = \mathbf{w}_{opt}$.

Consequently, the Lagrangian for this problem can be set up as

$$J_1(\mathbf{w}_k, \lambda) = J(\mathbf{w}_k) + \lambda\left[J(\mathbf{w}_k) - J_{\min}\right] \tag{4}$$

The critical values of Eq. (4) are $\mathbf{w}_k = \mathbf{w}_{opt}$ for any $\gamma$. This situation may cause convergence problem. So, a new augmented Lagrangian as follows:

$$J_2(\mathbf{w}_k, \lambda) = J(\mathbf{w}_k) + \gamma\lambda\left[J(\mathbf{w}_k) - J_{\min}\right] - \gamma\lambda^2 \tag{5}$$

The uses of the $\gamma$ and $\gamma\lambda^2$ in Eq. (5) ensure the unique critical value of $J_2(\mathbf{w}_k, \lambda)$ is $(\mathbf{w}_k, \lambda) = (\mathbf{w}_{opt}, 0)$. In Eq. (5), the augmented Lagrangian is minimized with respect to the weight $\mathbf{w}_k$ and maximized with respect to $\lambda$, respectively. By applying the Robbnis-Munro method [3], the weight and $\lambda$ are updated as follows:

$$\mathbf{w}_{k+1} = \mathbf{w}_k - \alpha \nabla_{\mathbf{w}} J_2 \left( \mathbf{w}_k, \lambda \right) \tag{6}$$

$$\lambda_{k+1} = \lambda_k + \beta \nabla_{\lambda} J_2 \left( \mathbf{w}_k, \lambda \right) \tag{7}$$

where $\alpha$ and $\beta$ are the positive learning parameters.

The cost function to be minimized is given by

$$J(\mathbf{w}) = \mathrm{E}\left[ |e_k|^3 \right] \tag{8}$$

Substituting the results for derivatives in Eq. (6) and Eq. (7), and redefining $\alpha$ and $\beta$ (replace $\alpha$ by $\alpha/3$ and $\beta\gamma$ by $\beta/2$), we obtain the following update recursions:

$$\mathbf{w}_{k+1} = \mathbf{w}_k + \alpha_k e_k^2 \operatorname{sgn}(e_k) \mathbf{x}_k \tag{9}$$

$$\alpha_k = \alpha \left( 1 + \gamma \lambda_k \right) \tag{10}$$

$$\lambda_{k+1} = (1 - \beta) \lambda_k + \frac{\beta}{2} \left( |e_k|^3 - J_{\min} \right) \tag{11}$$

where $\operatorname{sgn}(e_k)$ denotes the sign function of the variable $e_k$, i.e., if $e_k \geq 0$, then $\operatorname{sgn}(e_k) = 1$, otherwise $\operatorname{sgn}(e_k) = -1$. $J_{\min} = \sigma_\xi^2$. $\alpha$, $\beta$ and $\gamma$ are positive parameters providing more control over the performance of the proposed algorithm. As suggested in [3], after fixing the nominal step-size $\alpha$, $\gamma$ should be chosen as a value as large as possible to obtain a fast convergence rate. In additions, $\beta$ should be chosen as a small value to achieve a desired MSE and MSD.

If $J_{\min} = 0$ in (11) as, $\lambda_{k+1} = (1 - \beta) \lambda_k + \frac{\beta}{2} |e_k|^3$ ,we get what we call the zero noise constrained LMAT (ZNCLMAT) algorithm, which is similar to the NCLMAT but results in a mismatch due to the fact that the true noise absolute third moment is replaced by zero. The ZNCLMAT can be analyzed using the same approach used to analyze the NCLMAT in the sequel. Also, it can be observed from Eq. (9) and Eq. (10) that the LMAT algorithm [10] is recovered when $\gamma$=0.

## 3. Performances of the NCLMAT algorithm

In this section, the mean behavior and the second order moments of the proposed NCLMAT algorithm are analyzed. We make the following assumptions to analyze the NCLMAT algorithm [9][12]:

*A1:* $\{\mathbf{x}_k\}$ *is a wide-sense stationary sequence of i.i.d. Gaussian random variables with zero-mean and positive definite autocorrelation matrix* $\mathbf{R}_{\mathbf{xx}}$.

*A2:* $\{\xi_k\}$ *is independent of the input process.*

*A3: For any time instant k,* $\alpha_k$ *and* $\mathbf{w}_k$ *are statistically independent.*

*A4: For any time instant k,* $\mathbf{x}_k$ *and* $\mathbf{v}_k = \mathbf{w}_{opt} - \mathbf{w}_k$ *are statistically independent.*

*A5:* $e(n)$ *conditioned on* $\mathbf{v}_k$ *is Gaussian with zero mean and* $\sigma_{e|\mathbf{v}}^2(k) \approx \sigma_e^2(k)$.

*A6: The statistical dependence of* $\mathbf{v}_k \mathbf{v}_k^{\mathrm{T}}$ *and* $\mathbf{v}_k$ *can be neglected.*

*A7:* $\mathbf{v}_k$ *and* $\mathbf{x}_k^{\mathrm{T}} \mathbf{v}_k$ *are statistically independent.*

Note that when parameters are chosen so that the steady-state variance of $\alpha_k$ or $\mathbf{w}_k$ is small, then *A4* can be approximately satisfied.

### *3.1. The computation complexity*

Unlike the LMAT algorithm, the parameters $\alpha_k$ and $\lambda_k$ required to calculate in NCLMAT algorithm. So, computational complexity of the NCLMAT algorithm is much than the LMAT algorithm. Furthermore, there is no $\operatorname{sgn}(\bullet)$ required to operate in NCLMF algorithm. But there is a higher order power operation for the error in NCLMF algorithm. For convenience, the computational complexity of the NCLMAT algorithm and that of other existing algorithms is listed in Table 1. As can be seen from Table 1, the computational complexity of NCLMF algorithm and NCLMAT algorithm is approximate.

Table 1. The computational complexity of NCLMAT, LMAT and NCLMF algorithms

| Algorithm | Multiplications | Comparisons | Additions | $(\ )^2$ | $(\ )^3$ |
|---|---|---|---|---|---|
| LMAT | 2*N*+2 | 1 | 2*N*+1 | 1 | 0 |
| NCLMF | 2*N*+9 | 0 | 2*N*+5 | 2 | 1 |
| NCLMAT | 2*N*+7 | 1 | 2*N*+5 | 1 | 1 |

### *3.2. Mean weight behavior model*

Based on Eq. (9), defining the weight error vector,

$$\mathbf{v}_{k+1} = \mathbf{v}_k - \alpha_k e_k^2 \operatorname{sgn}(e_k)\mathbf{x}_k \tag{12}$$

Expression Eq. (11) can be rewritten as

$$\begin{aligned}
&\lambda_{k+1} \\
&= (1-\beta)\lambda_k + \frac{\beta}{2}\left(|e_k|^3 - J_{\min}\right) \\
&= (1-\beta)^2 \lambda_{k-1} + \frac{\beta}{2}(1-\beta)\left(|e_{k-1}|^3 - J_{\min}\right) + \frac{\beta}{2}\left(|e_k|^3 - J_{\min}\right) \\
&= (1-\beta)^3 \lambda_{k-2} + \frac{\beta}{2}(1-\beta)^2\left(|e_{k-2}|^3 - J_{\min}\right) + \frac{\beta}{2}(1-\beta)\left(|e_{k-1}|^3 - J_{\min}\right) + \frac{\beta}{2}\left(|e_k|^3 - J_{\min}\right) \\
&\cdots \\
&= (1-\beta)^{k+1}\lambda_0 + \frac{\beta}{2}\sum_{n=0}^{k}\left[(1-\beta)^n\left(|e_{k-n}|^3 - J_{\min}\right)\right]
\end{aligned} \tag{13}$$

$\lambda_0$ has an initial value of 0.

So

$$\lambda_k = \frac{\beta}{2}\sum_{n=0}^{k-1}\left[(1-\beta)^{k-1-n}\left(|e_n|^3 - J_{\min}\right)\right] \tag{14}$$

Substituting Eq. (14) into Eq. (10),

$$\alpha_k = \alpha + \frac{\alpha\beta\gamma}{2}\sum_{n=0}^{k-1}\left[(1-\beta)^{k-1-n}\left(|e_n|^3 - J_{\min}\right)\right] \tag{15}$$

Then, defining $\delta = \alpha\gamma\beta/2$, we get

$$\alpha_k = \alpha + \delta\sum_{n=0}^{k-1}\left[(1-\beta)^{k-1-n}\left(|e_n|^3 - J_{\min}\right)\right] \tag{16}$$

Now, substituting Eq. (16) into Eq. (12),

$$\begin{aligned}
&\mathbf{v}_{k+1} \\
&= \mathbf{v}_k - \left\{ \alpha + \delta \sum_{n=0}^{k-1} \left[ (1-\beta)^{k-1-n} \left( |e_n|^3 - J_{\min} \right) \right] \right\} e_k^2 \operatorname{sgn}(e_k) \mathbf{x}_k \\
&= \mathbf{v}_k - \alpha e_k^2 \operatorname{sgn}(e_k) \mathbf{x}_k - \delta \sum_{n=0}^{k-1} \left[ (1-\beta)^{k-1-n} \left( |e_n|^3 - J_{\min} \right) \right] e_k^2 \operatorname{sgn}(e_k) \mathbf{x}_k
\end{aligned} \tag{17}$$

Redefining the time-varying term as

$$\mu_k = \delta \sum_{n=0}^{k-1} \left[ (1-\beta)^{k-1-n} \left( |e_n|^3 - J_{\min} \right) \right] \tag{18}$$

We obtain the final expression for the weight error vector update equation

$$\mathbf{v}_{k+1} = \mathbf{v}_k - \alpha e_k^2 \operatorname{sgn}(e_k) \mathbf{x}_k - \mu_k e_k^2 \operatorname{sgn}(e_k) \mathbf{x}_k \tag{19}$$

Using the above mentioned assumptions, it can be shown that the mean behavior of the weight error vector is governed by the following recursion,

$$\mathrm{E}[\mathbf{v}_{k+1}] = \mathrm{E}[\mathbf{v}_k] - \alpha \, \mathrm{E}\left[ e_k^2 \operatorname{sgn}(e_k) \mathbf{x}_k \right] - \mathrm{E}\left[ \mu_k e_k^2 \operatorname{sgn}(e_k) \mathbf{x}_k \right] \tag{20}$$

Note that Eq. (20) is identical to the mean weight value expression of the standard LMAT algorithm except for the last term, which can be written as follows,

$$\begin{aligned}
&\mathrm{E}\left[ \mu_k e_k^2 \operatorname{sgn}(e_k) \mathbf{x}_k \right] \\
&= \delta \, \mathrm{E}\left\{ \sum_{n=0}^{k-1} \left[ (1-\beta)^{k-1-n} |e_n|^3 \right] e_k^2 \operatorname{sgn}(e_k) \mathbf{x}_k \right\} - \delta \, \mathrm{E}\left[ \sum_{n=0}^{k-1} (1-\beta)^{k-1-n} J_{\min} e_k^2 \operatorname{sgn}(e_k) \mathbf{x}_k \right] \\
&= \delta \, \mathrm{E}\left\{ \sum_{n=0}^{k-1} \left[ (1-\beta)^{k-1-n} |e_n|^3 \right] e_k^2 \operatorname{sgn}(e_k) \mathbf{x}_k \right\} - \delta J_{\min} \sum_{n=0}^{k-1} (1-\beta)^{k-1-n} \, \mathrm{E}\left[ e_k^2 \operatorname{sgn}(e_k) \mathbf{x}_k \right]
\end{aligned} \tag{21}$$

We readily note that $\sum_{n=0}^{k-1} (1-\beta)^{k-1-n} |e_n|^3$ is statistically independent of resulting in the following relation,

$$\delta \, \mathrm{E}\left\{ \sum_{n=0}^{k-1} \left[ (1-\beta)^{k-1-n} |e_n|^3 \right] e_k^2 \operatorname{sgn}(e_k) \mathbf{x}_k \right\} = \delta \, \mathrm{E}\left\{ \sum_{n=0}^{k-1} \left[ (1-\beta)^{k-1-n} |e_n|^3 \right] \right\} \mathrm{E}\left[ e_k^2 \operatorname{sgn}(e_k) \mathbf{x}_k \right] \tag{22}$$

Substituting Eq. (21) and Eq. (22) into Eq. (20), we get

$$\begin{aligned}
&\mathrm{E}[\mathbf{v}_{k+1}] \\
&= \mathrm{E}[\mathbf{v}_k] - \alpha \, \mathrm{E}\left[ e_k^2 \operatorname{sgn}(e_k) \mathbf{x}_k \right] - \mathrm{E}\left[ \mu_k e_k^2 \operatorname{sgn}(e_k) \mathbf{x}_k \right] \\
&= \mathrm{E}[\mathbf{v}_k] - \left\{ \alpha + \delta \, \mathrm{E}\left\{ \sum_{n=0}^{k-1} \left[ (1-\beta)^{k-1-n} |e_n|^3 \right] \right\} - \delta J_{\min} \sum_{n=0}^{k-1} (1-\beta)^{k-1-n} \right\} \mathrm{E}\left[ e_k^2 \operatorname{sgn}(e_k) \mathbf{x}_k \right]
\end{aligned} \tag{23}$$

The distribution function of $Y = e_k^2$ conditioned on $\mathbf{v}_k$ is $F(y)$ under *A5*,

$$\begin{aligned}
&F(y) \\
&= p\left( e_k^2 < y \right) \\
&\approx \int_{-\sqrt{y}}^{\sqrt{y}} \frac{1}{\sqrt{2\pi}\sigma_{e,k}} \exp\left(-\frac{e_k^2}{2\sigma_{e,k}^2}\right) \mathrm{d}\, e_k
\end{aligned} \tag{24}$$

then the probability density function of $Y = e_k^2$ conditioned on $\mathbf{v}_k$ is shown as Eq. (25),

$$\begin{aligned}
&p(y) \\
&= \nabla_y \left[ F(y) \right] \\
&= \frac{1}{\sqrt{2\pi y}\sigma_{e,k}} \exp\left(-\frac{y}{2\sigma_{e,k}^2}\right)
\end{aligned} \tag{25}$$

The distribution function of $Y = |e_k|$ conditioned on $\mathbf{v}_k$ is $F(y)$ under *A5*,

$$\begin{aligned} &F(y) \\ &= p\left(|e_k| < y\right) \\ &\approx \int_{-y}^{y} \frac{1}{\sqrt{2\pi}\sigma_{e,k}} \exp(-\frac{e_k^2}{2\sigma_{e,k}^2}) \,\mathrm{d}\, e_k \end{aligned} \tag{26}$$

then the probability density function of $Y = |e_k|$ conditioned on $\mathbf{v}_k$ is shown as Eq. (27),

$$\begin{aligned} &p(y) \\ &= \nabla_y \left[F(y)\right] \\ &= \frac{2}{\sqrt{2\pi}\sigma_{e,k}} \exp(-\frac{y^2}{2\sigma_{e,k}^2}) \end{aligned} \tag{27}$$

So,

$$\mathrm{E}\left[|e_k|\right] = \sqrt{\frac{2}{\pi}}\sigma_{e,k} \tag{28}$$

and

$$\mathrm{E}\left[|e_k|^3\right] = 2\sqrt{\frac{2}{\pi}}\sigma_{e,k}^3 \tag{29}$$

and

$$\begin{aligned} &\mathrm{E}\left[\mathbf{x}_k e_k^2 \operatorname{sgn}(e_k)\right] \\ &= \mathrm{E}\left\{\mathrm{E}\left[\mathbf{x}_k e_k^2 \operatorname{sgn}(e_k) \mid \mathbf{v}_k\right]\right\} \\ &= 2\sqrt{\frac{2}{\pi}}\sigma_{e,k}\mathbf{R}_{\mathbf{xx}}\,\mathrm{E}\left[\mathbf{v}_k\right] \end{aligned} \tag{30}$$

So,

$$\begin{aligned} &\mathrm{E}\left[\mu_k\right] \\ &= \delta\sum_{n=0}^{k-1}\left[(1-\beta)^{k-1-n}\left(\mathrm{E}\left[|e_n|^3\right] - J_{\min}\right)\right] \\ &= \delta\sum_{n=0}^{k-1}\left[(1-\beta)^{k-1-n}\left(2\sqrt{\frac{2}{\pi}}\sigma_{e,n}^3 - J_{\min}\right)\right] \\ &= \delta\sum_{n=0}^{k-1}(1-\beta)^{k-1-n}\sum_{n=0}^{k-1}\left(2\sqrt{\frac{2}{\pi}}\sigma_{e,n}^3 - J_{\min}\right) \end{aligned} \tag{31}$$

Thus,

$$\begin{aligned} &\mathrm{E}\left[\mathbf{v}_{k+1}\right] \\ &= \mathrm{E}\left[\mathbf{v}_k\right] - \left\{\alpha + \delta\,\mathrm{E}\left[\sum_{n=0}^{k-1}(1-\beta)^{k-1-n}|e_n|^3\right] - \delta J_{\min}\sum_{n=0}^{k-1}(1-\beta)^{k-1-n}\right\}\mathrm{E}\left[e_k^2 \operatorname{sgn}\left(e_k\right)\mathbf{x}_k\right] \\ &= \mathrm{E}\left[\mathbf{v}_k\right] - \left\{\alpha + \delta\left[\sum_{n=0}^{k-1}(1-\beta)^{k-1-n}\,\mathrm{E}\left[|e_n|^3\right]\right] - \delta J_{\min}\sum_{n=0}^{k-1}(1-\beta)^{k-1-n}\right\}2\sqrt{\frac{2}{\pi}}\sigma_{e,k}\mathbf{R}_{\mathbf{xx}}\,\mathrm{E}\left[\mathbf{v}_k\right] \\ &= \mathrm{E}\left[\mathbf{v}_k\right] - \sqrt{\frac{2}{\pi}}\alpha\left\{2 + 2\sqrt{\frac{2}{\pi}}\sum_{n=0}^{k-1}\sigma_{e,n}^3\gamma\left(1-(1-\beta)^k\right) - \gamma J_{\min}\left(1-(1-\beta)^k\right)\right\}\sigma_{e,k}\mathbf{R}_{\mathbf{xx}}\,\mathrm{E}\left[\mathbf{v}_k\right] \\ &= \left\{\mathbf{I} - \sqrt{\frac{2}{\pi}}\alpha\left[2 + \gamma\left(2\sqrt{\frac{2}{\pi}}\sum_{n=0}^{k-1}\sigma_{e,n}^3 - J_{\min}\right)\left(1-(1-\beta)^k\right)\right]\sigma_{e,k}\mathbf{R}_{\mathbf{xx}}\right\}\mathrm{E}\left[\mathbf{v}_k\right] \end{aligned} \tag{32}$$

Condition for the stability of the mean weight error vector (as Eq. (32)) is given by

$$0 < \alpha < \frac{2}{\sqrt{\frac{2}{\pi}}\left[2+\gamma\left(2\sqrt{\frac{2}{\pi}}\sum_{n=0}^{k-1}\sigma_{e,n}^{3} - J_{\min}\right)\left(1-(1-\beta)^{k}\right)\right]\sigma_{e,k}\rho_{\max}} \tag{33}$$

where $\rho_{\max}$ is the maximum eigenvalue associated with $\mathbf{R}_{\mathbf{xx}}$. In addition, when $\gamma = 0$, this range of $\alpha$ is the LMAT algorithm. As seen from Eq. (32), we will get $\mathrm{E}[\mathbf{v}_\infty] = 0$ as $k \to \infty$.

*3.3. Second-order moment analysis*

The MSE is given by

$$\begin{aligned} &\mathrm{E}\left[e_k^2\right] \\ &= \mathrm{E}\left[\left(\mathbf{v}_k^{\mathrm{T}}\mathbf{x}_k + \xi_k\right)\left(\mathbf{x}_k^{\mathrm{T}}\mathbf{v}_k + \xi_k\right)\right] \\ &= \sigma_\xi^2 + \mathrm{E}\left[\mathbf{v}_k^{\mathrm{T}}\mathbf{x}_k\mathbf{x}_k^{\mathrm{T}}\mathbf{v}_k\right] \\ &= \sigma_\xi^2 + \mathrm{tr}\left\{\mathbf{R}_{\mathbf{xx}}\,\mathrm{E}\left[\mathbf{v}_k\mathbf{v}_k^{\mathrm{T}}\right]\right\} \\ &= \sigma_\xi^2 + \sigma_x^2\,\mathrm{MSD}_k \end{aligned} \tag{34}$$

where $\mathrm{MSD}_k = \mathrm{E}\left[\|\mathbf{v}_k\|^2\right]$.

Pre-multiplying Eq. (12) by its transpose form yields,

$$\begin{aligned} &\|\mathbf{v}_{k+1}\|^2 \\ &= \left[\mathbf{v}_k^{\mathrm{T}} - \alpha e_k^2\,\mathrm{sgn}(e_k)\mathbf{x}_k^{\mathrm{T}} - \mu_k e_k^2\,\mathrm{sgn}(e_k)\mathbf{x}_k^{\mathrm{T}}\right]\left[\mathbf{v}_k - \alpha e_k^2\,\mathrm{sgn}(e_k)\mathbf{x}_k - \mu_k e_k^2\,\mathrm{sgn}(e_k)\mathbf{x}_k\right] \\ &= \mathbf{v}_k^{\mathrm{T}}\mathbf{v}_k - \alpha e_k^2\,\mathrm{sgn}(e_k)\mathbf{v}_k^{\mathrm{T}}\mathbf{x}_k - \mu_k e_k^2\,\mathrm{sgn}(e_k)\mathbf{v}_k^{\mathrm{T}}\mathbf{x}_k - \alpha e_k^2\,\mathrm{sgn}(e_k)\mathbf{x}_k^{\mathrm{T}}\mathbf{v}_k - \mu_k e_k^2\,\mathrm{sgn}(e_k)\mathbf{x}_k^{\mathrm{T}}\mathbf{v}_k \\ &\quad + \alpha^2 e_k^4\mathbf{x}_k^{\mathrm{T}}\mathbf{x}_k + \alpha\mu_k e_k^4\mathbf{x}_k^{\mathrm{T}}\mathbf{x}_k + \alpha\mu_k e_k^4\mathbf{x}_k^{\mathrm{T}}\mathbf{x}_k + \mu_k^2 e_k^4\mathbf{x}_k^{\mathrm{T}}\mathbf{x}_k \end{aligned} \tag{35}$$

Taking the expected value, and we also assume that $N$ is sufficiently large so that $\mathrm{E}\left[\mathbf{x}_k^{\mathrm{T}}\mathbf{x}_k\right] \approx N\sigma_x^2$ because of an essentially ergodic assumption such that the time average over the taps is equal to the ensemble average.

$$\begin{aligned} &\mathrm{E}\left[\|\mathbf{v}_{k+1}\|^2\right] \\ &= \mathrm{E}\left[\|\mathbf{v}_k\|^2\right] - \alpha\,\mathrm{E}\left[e_k^2\,\mathrm{sgn}(e_k)\mathbf{v}_k^{\mathrm{T}}\mathbf{x}_k\right] - \mathrm{E}\left[\mu_k e_k^2\,\mathrm{sgn}(e_k)\mathbf{v}_k^{\mathrm{T}}\mathbf{x}_k\right] - \alpha\,\mathrm{E}\left[e_k^2\,\mathrm{sgn}(e_k)\mathbf{x}_k^{\mathrm{T}}\mathbf{v}_k\right] \\ &\qquad - \mathrm{E}\left[\mu_k e_k^2\,\mathrm{sgn}(e_k)\mathbf{x}_k^{\mathrm{T}}\mathbf{v}_k\right] + \alpha^2\,\mathrm{E}\left[e_k^4\mathbf{x}_k^{\mathrm{T}}\mathbf{x}_k\right] + 2\alpha\,\mathrm{E}\left[\mu_k e_k^4\mathbf{x}_k^{\mathrm{T}}\mathbf{x}_k\right] + \mathrm{E}\left[\mu_k^2 e_k^4\mathbf{x}_k^{\mathrm{T}}\mathbf{x}_k\right] \\ &= \mathrm{E}\left[\|\mathbf{v}_k\|^2\right] - 2\left(\alpha + \mathrm{E}[\mu_k]\right)\mathrm{E}\left[e_k^2\,\mathrm{sgn}(e_k)\mathbf{v}_k^{\mathrm{T}}\mathbf{x}_k\right] + \left(\alpha^2 + 2\alpha\,\mathrm{E}[\mu_k] + \mathrm{E}\left[\mu_k^2\right]\right)\mathrm{E}\left[e_k^4\mathbf{x}_k^{\mathrm{T}}\mathbf{x}_k\right] \\ &= \mathrm{E}\left[\|\mathbf{v}_k\|^2\right] - 2\left(\alpha + \mathrm{E}[\mu_k]\right)2\sqrt{\frac{2}{\pi}}\sigma_{e,k}\,\mathrm{E}\left[\left(\mathbf{v}_k^{\mathrm{T}}\mathbf{x}_k\right)^2\right] + \left(\alpha^2 + 2\alpha\,\mathrm{E}[\mu_k] + \mathrm{E}\left[\mu_k^2\right]\right)\mathrm{E}\left[e_k^4\mathbf{x}_k^{\mathrm{T}}\mathbf{x}_k\right] \end{aligned} \tag{36}$$

To further simplify Eq. (36), we can take advantage of Eq. (13) and Eq. (17) in [11], i.e.

$$\mathrm{E}\left[\left(\mathbf{v}_k^{\mathrm{T}}\mathbf{x}_k\right)^2\right] = \sigma_x^2\,\mathrm{E}\left[\mathbf{v}_k^{\mathrm{T}}\mathbf{v}_k\right] \tag{37}$$

$$
\begin{aligned}
&\mathrm{E}\left[e_k^4\mathbf{x}_k^{\mathrm{T}}\mathbf{x}_k\right]\\
&=\mathrm{E}\left[\left(\xi_k+\mathbf{v}_k^{\mathrm{T}}\mathbf{x}_k\right)^4\mathbf{x}_k^{\mathrm{T}}\mathbf{x}_k\right]\\
&=\mathrm{E}\left\{\left[\xi_k^4+6\xi_k^2\left(\mathbf{v}_k^{\mathrm{T}}\mathbf{x}_k\right)^2+\left(\mathbf{v}_k^{\mathrm{T}}\mathbf{x}_k\right)^4\right]\mathbf{x}_k^{\mathrm{T}}\mathbf{x}_k\right\}\\
&=\mathrm{E}\left[\xi_k^4\mathbf{x}_k^{\mathrm{T}}\mathbf{x}_k\right]+6\sigma_\xi^2\,\mathrm{E}\left[\left(\mathbf{v}_k^{\mathrm{T}}\mathbf{x}_k\right)^2\mathbf{x}_k^{\mathrm{T}}\mathbf{x}_k\right]+\mathrm{E}\left[\left(\mathbf{v}_k^{\mathrm{T}}\mathbf{x}_k\right)^4\mathbf{x}_k^{\mathrm{T}}\mathbf{x}_k\right]\\
&\approx N\sigma_x^2\,\mathrm{E}\left[\xi_k^4\right]+6(N+2)\sigma_\xi^2\sigma_x^4\,\mathrm{E}\left[\mathbf{v}_k^{\mathrm{T}}\mathbf{v}_k\right]+3(N+4)\sigma_x^6\,\mathrm{E}^2\left[\mathbf{v}_k^{\mathrm{T}}\mathbf{v}_k\right]
\end{aligned}
\tag{38}
$$

Substituting Eq. (37) and Eq. (38) into Eq. (36),

$$
\begin{aligned}
&\mathrm{E}\left[\left\|\mathbf{v}_{k+1}\right\|^2\right]\\
&=\mathrm{E}\left[\left\|\mathbf{v}_k\right\|^2\right]-2\left(\alpha+\mathrm{E}\left[\mu_k\right]\right)2\sqrt{\frac{2}{\pi}}\sigma_{e,k}\,\mathrm{E}\left[\left(\mathbf{v}_k^{\mathrm{T}}\mathbf{x}_k\right)^2\right]+\left(\alpha^2+2\alpha\,\mathrm{E}\left[\mu_k\right]+\mathrm{E}\left[\mu_k^2\right]\right)\mathrm{E}\left[e_k^4\mathbf{x}_k^{\mathrm{T}}\mathbf{x}_k\right]\\
&=\mathrm{E}\left[\left\|\mathbf{v}_k\right\|^2\right]-4\sqrt{\frac{2}{\pi}}\left(\alpha+\mathrm{E}\left[\mu_k\right]\right)\sigma_{e,k}\sigma_x^2\,\mathrm{E}\left[\left\|\mathbf{v}_k\right\|^2\right]\\
&\quad+\left(\alpha^2+2\alpha\,\mathrm{E}\left[\mu_k\right]+\mathrm{E}\left[\mu_k^2\right]\right)\left[N\sigma_x^2\,\mathrm{E}\left[\xi_k^4\right]+6(N+2)\sigma_\xi^2\sigma_x^4\,\mathrm{E}\left[\left\|\mathbf{v}_k\right\|^2\right]+3(N+4)\sigma_x^6\,\mathrm{E}^2\left[\left\|\mathbf{v}_k\right\|^2\right]\right]
\end{aligned}
\tag{39}
$$

So

$$
\begin{aligned}
&\mathrm{MSD}_{k+1}\\
&=\mathrm{MSD}_k-4\sqrt{\frac{2}{\pi}}\left(\alpha+\mathrm{E}\left[\mu_k\right]\right)\sigma_{e,k}\sigma_x^2\,\mathrm{MSD}_k\\
&\quad+\left(\alpha^2+2\alpha\,\mathrm{E}\left[\mu_k\right]+\mathrm{E}\left[\mu_k^2\right]\right)\left[N\sigma_x^2\,\mathrm{E}\left[\xi_k^4\right]+6(N+2)\sigma_\xi^2\sigma_x^4\,\mathrm{MSD}_k+3(N+4)\sigma_x^6\,\mathrm{MSD}_k^2\right]
\end{aligned}
\tag{40}
$$

Eq. (40) will express as Eq. (41) when discard $\mathrm{MSD}_k^2$.

$$
\begin{aligned}
&\mathrm{MSD}_{k+1}\\
&=\mathrm{MSD}_k-4\sqrt{\frac{2}{\pi}}\left(\alpha+\mathrm{E}\left[\mu_k\right]\right)\sigma_{e,k}\sigma_x^2\,\mathrm{MSD}_k+6(N+2)\sigma_\xi^2\sigma_x^4\left(\alpha^2+2\alpha\,\mathrm{E}\left[\mu_k\right]+\mathrm{E}\left[\mu_k^2\right]\right)\mathrm{MSD}_k\\
&\quad+N\sigma_x^2\left(\alpha^2+2\alpha\,\mathrm{E}\left[\mu_k\right]+\mathrm{E}\left[\mu_k^2\right]\right)\mathrm{E}\left[\xi_k^4\right]
\end{aligned}
\tag{41}
$$

To further simplify Eq. (46), we can take advantage of Eq. (31), Eq. (42), Eq. (43) and Eq. (44), i.e.

$$
\begin{cases}
\displaystyle\sum_{n=0}^{k-1}(1-\beta)^{k-1-n}=\frac{1-(1-\beta)^k}{\beta}\\
\displaystyle\sum_{n=0}^{k-1}(1-\beta)^{2(k-1-n)}=\frac{1-(1-\beta)^{2k}}{\beta(2-\beta)}
\end{cases}
\tag{42}
$$

and

$$
\begin{aligned}
&\mathrm{E}\left[e_k^6\right]\\
&=\frac{2}{\sqrt{2\pi}\sigma_{e,k}}\int_0^{+\infty}y^6\exp\left(-\frac{y^2}{2\sigma_{e,k}^2}\right)\mathrm{d}\,y\\
&=15\sigma_{e,k}^6
\end{aligned}
\tag{43}
$$

and

$$
\begin{aligned}
&\mathrm{E}\left[\mu_k^2\right] \\
&= \delta^2 \sum_{n=0}^{k-1} (1-\beta)^{2(k-1-n)} \mathrm{E}\left[\left(\left|e_n\right|^3 - J_{\min}\right)^2\right] \\
&= \delta^2 \sum_{n=0}^{k-1} (1-\beta)^{2(k-1-n)} \left(\mathrm{E}\left[e_n^6\right] - 2J_{\min}\mathrm{E}\left[\left|e_n\right|^3\right] + J_{\min}^2\right) \\
&= \delta^2 \sum_{n=0}^{k-1} \left[(1-\beta)^{2(k-1-n)} \left(15\sigma_{e,n}^6 - 4\sqrt{\frac{2}{\pi}}\sigma_{e,n}^3 J_{\min} + J_{\min}^2\right)\right]
\end{aligned}
\tag{47}
$$

So

$$
\begin{aligned}
&\mathrm{MSD}_{k+1} \\
&= \mathrm{MSD}_k - 2\sqrt{\frac{2}{\pi}}\alpha\left[2 + \gamma A(k)\left(2\sqrt{\frac{2}{\pi}}\sum_{n=0}^{k-1}\sigma_{e,n}^3 - \sum_{n=0}^{k-1} J_{\min}\right)\right]\sigma_{e,k}\sigma_x^2 \mathrm{MSD}_k \\
&\quad + 6(N+2)\alpha^2\sigma_\xi^2\sigma_x^4\left[1 + \gamma A(k)\left(2\sqrt{\frac{2}{\pi}}\sum_{n=0}^{k-1}\sigma_{e,n}^3 - \sum_{n=0}^{k-1} J_{\min}\right) + \frac{\beta\gamma^2 A(k)B(k)}{4(2-\beta)}\left(15\sum_{n=0}^{k-1}\sigma_{e,n}^6 - 4\sqrt{\frac{2}{\pi}}J_{\min}\sum_{n=0}^{k-1}\sigma_{e,n}^3 + \sum_{n=0}^{k-1}J_{\min}^2\right)\right]\mathrm{MSD}_k \\
&\quad + N\alpha^2\sigma_x^2\left[1 + \gamma A(k)\left(2\sqrt{\frac{2}{\pi}}\sum_{n=0}^{k-1}\sigma_{e,n}^3 - \sum_{n=0}^{k-1} J_{\min}\right) + \frac{\beta\gamma^2 A(k)B(k)}{4(2-\beta)}\left(15\sum_{n=0}^{k-1}\sigma_{e,n}^6 - 4\sqrt{\frac{2}{\pi}}J_{\min}\sum_{n=0}^{k-1}\sigma_{e,n}^3 + \sum_{n=0}^{k-1}J_{\min}^2\right)\right]\mathrm{E}\left[\xi_k^4\right]
\end{aligned}
\tag{48}
$$

where $A(k) = \left(1-(1-\beta)^k\right)$ and $B(k) = \left(1+(1-\beta)^k\right)$. As $\beta$ should be chosen as a small value to achieve a desired MSE and MSD, in addition, according to the existing literature [3][6], we can know $A(\infty) = 1$ and $B(\infty) = 1$ when $0 < \beta < 2,\ k \to \infty$.

Eq. (48) will express as Eq. (49) when discard $\mathrm{MSD}_k^t$, for $t \geq 2$.

$$
\begin{aligned}
&\mathrm{MSD}_{k+1} \\
&= \mathrm{MSD}_k - 2\sqrt{\frac{2}{\pi}}\alpha\left[2 + \gamma A(k)\left(2\sqrt{\frac{2}{\pi}}\sum_{n=0}^{k-1}\sigma_{e,n}^3 - \sum_{n=0}^{k-1} J_{\min}\right)\right]\sigma_{e,k}\sigma_x^2 \mathrm{MSD}_k \\
&\quad + 6(N+2)\alpha^2\sigma_\xi^2\sigma_x^4\left[1 + \gamma A(k)\left(2\sqrt{\frac{2}{\pi}}\sum_{n=0}^{k-1}\sigma_{e,n}^3 - \sum_{n=0}^{k-1} J_{\min}\right) + \frac{\beta\gamma^2 A(k)B(k)}{4(2-\beta)}\left(15\sum_{n=0}^{k-1}\sigma_{e,n}^6 - 4\sqrt{\frac{2}{\pi}}J_{\min}\sum_{n=0}^{k-1}\sigma_{e,n}^3 + \sum_{n=0}^{k-1}J_{\min}^2\right)\right]\mathrm{MSD}_k \\
&\quad + N\alpha^2\sigma_x^2\left[1 + \gamma A(k)\left(2\sqrt{\frac{2}{\pi}}\sum_{n=0}^{k-1}\sigma_{e,n}^3 - \sum_{n=0}^{k-1} J_{\min}\right) + \frac{\beta\gamma^2 A(k)B(k)}{4(2-\beta)}\left(15\sum_{n=0}^{k-1}\sigma_{e,n}^6 - 4\sqrt{\frac{2}{\pi}}J_{\min}\sum_{n=0}^{k-1}\sigma_{e,n}^3 + \sum_{n=0}^{k-1}J_{\min}^2\right)\right]\mathrm{E}\left[\xi_k^4\right]
\end{aligned}
\tag{49}
$$

So

$$
\begin{aligned}
&\mathrm{MSD}_{k+1} \\
&= \mathrm{MSD}_k - 2\sqrt{\frac{2}{\pi}}\alpha\left[2 + \gamma A(k)D(k)\right]\sigma_{e,k}\sigma_x^2\mathrm{MSD}_k \\
&\quad + 6(N+2)\alpha^2\sigma_\xi^2\sigma_x^4\left[1 + \gamma A(k)D(k) + \frac{\beta\gamma^2 A(k)B(k)F(k)}{4(2-\beta)}\right]\mathrm{MSD}_k \\
&\quad + N\alpha^2\sigma_x^2\left[1 + \gamma A(k)D(k) + \frac{\beta\gamma^2 A(k)B(k)F(k)}{4(2-\beta)}\right]\mathrm{E}\left[\xi_k^4\right]
\end{aligned}
\tag{50}
$$

where $C(k) = 2\sqrt{\frac{2}{\pi}}\sum_{n=0}^{k-1}\sigma_{e,n}^3$, $D(k) = C(k) - \sum_{n=0}^{k-1}J_{\min}$ and $F(k) = 15\sum_{n=0}^{k-1}\sigma_{e,n}^6 + \sum_{n=0}^{k-1}J_{\min}^2 - 2J_{\min}C(k)$.

Therefore, we obtain an expression of the MSD behavior of the NCLMAT algorithm, i.e.

$$
\mathrm{MSD}_{k+1} = f\left(N,\alpha,\beta,\gamma,\sigma_e,\sigma_x,\sigma_\xi\right)\mathrm{MSD}_k + g\left(N,\alpha,\beta,\gamma,\sigma_e,\sigma_x,\sigma_\xi\right) \tag{51}
$$

where

$$f\left(N,\alpha,\beta,\gamma,\sigma_e,\sigma_x,\sigma_\xi\right)$$
$$=1-2\sqrt{\frac{2}{\pi}}\alpha\left[2+\gamma A(k)D(k)\right]\sigma_{e,k}\sigma_x^2+6(N+2)\alpha^2\sigma_\xi^2\sigma_x^4\left[1+\gamma A(k)D(k)+\frac{\beta\gamma^2 A(k)B(k)F(k)}{4(2-\beta)}\right] \quad (52)$$

$$g\left(N,\alpha,\beta,\gamma,\sigma_e,\sigma_x,\sigma_\xi\right)=N\alpha^2\sigma_x^2\left[1+\gamma A(k)D(k)+\frac{\beta\gamma^2 A(k)B(k)F(k)}{4(2-\beta)}\right]\mathrm{E}\left[\xi_k^4\right] \quad (53)$$

The term $f\left(N,\alpha,\beta,\gamma,\sigma_e,\sigma_x,\sigma_\xi\right)$ influences the convergence rate of the algorithm. It can be noticed that the fastest convergence mode is obtained when the function from Eq. (52) reaches its minimum.

$$\alpha=\frac{\sqrt{\frac{2}{\pi}}\sigma_{e,k}\left[2+\gamma A(k)D(k)\right]}{6(N+2)\sigma_\xi^2\sigma_x^2\left[1+\gamma A(k)D(k)+\frac{\beta\gamma^2 A(k)B(k)F(k)}{4(2-\beta)}\right]} \quad (54)$$

For the NCLMAT algorithm to be stable in the mean-square sense, its MSD behavior must decrease monotonically with an increase of iteration $k$.

$$0<\alpha<\frac{\sqrt{\frac{2}{\pi}}\sigma_{e,k}\left[2+\gamma A(k)D(k)\right]}{2(N+2)\sigma_\xi^2\sigma_x^2\left[1+\gamma A(k)D(k)+\frac{\beta\gamma^2 A(k)B(k)F(k)}{4(2-\beta)}\right]} \quad (55)$$

Assuming that the NCLMAT algorithm has converged to the steady state,

$$\mathrm{MSD}_{k+1}=\mathrm{MSD}_k \quad (56)$$

The $\mathrm{MSD}_k$ is obtained by substituting Eq. (56) into Eq. (51) as follows

$$\mathrm{MSD}_k=\frac{N\alpha\left[N\gamma A(k)D(k)+\beta\gamma^2\frac{A(k)B(k)F(k)}{2-\beta}+1\right]\mathrm{E}\left[\xi_k^4\right]}{2\sqrt{\frac{2}{\pi}}\left[2+\gamma A(k)D(k)\right]\sigma_{e,k}-6(N+2)\alpha\sigma_\xi^2\sigma_x^2\left[1+\gamma A(k)D(k)+\frac{\beta\gamma^2 A(k)B(k)F(k)}{4(2-\beta)}\right]} \quad (57)$$

Finally, the $\mathrm{MSE}_k$ is given by incorporating Eq. (57) into Eq. (34),

$$\mathrm{MSE}_k=\sigma_\xi^2+\frac{N\alpha\sigma_x^2\left[N\gamma A(k)D(k)+\beta\gamma^2\frac{A(k)B(k)F(k)}{2-\beta}+1\right]\mathrm{E}\left[\xi_k^4\right]}{2\sqrt{\frac{2}{\pi}}\left[2+\gamma A(k)D(k)\right]\sigma_{e,k}-6(N+2)\alpha\sigma_\xi^2\sigma_x^2\left[1+\gamma A(k)D(k)+\frac{\beta\gamma^2 A(k)B(k)F(k)}{4(2-\beta)}\right]} \quad (58)$$

In addition, we can know $\mathrm{MSD}_\infty=0$ or $\mathrm{MSE}_\infty=\sigma_\xi^2$ when $a=0$.

In the real practical world $\sigma_e^2$ and $\sigma_x^2$ are unknown, however, we can estimate $\sigma_e^2$ and $\sigma_x^2$ by using equation Eq. (59) [13][14].

$$\begin{cases}\sigma_{e,k}^2=\dfrac{1}{\theta+\sigma_{x,k}^2}\mathbf{p}_k^{\mathrm{T}}\mathbf{p}_k\\ \sigma_{x,k}^2=\chi\sigma_{x,k}^2+(1-\chi)\mathbf{x}_k^{\mathrm{T}}\mathbf{x}_k\end{cases} \quad (59)$$

where $\mathbf{p}_k=\mathrm{E}\left[e_k\mathbf{x}_k\right]$, $\theta$ is a small positive number to avoid that the denominator of Eq. (59) is infinite when $\sigma_{x,k}^2=0$.

Although in Eq. (59), both $\mathbf{p}_k$ is also unknown, we can estimate this parameter by using Eq. (60).

$$\mathbf{p}_k=\chi\mathbf{p}_k+(1-\chi)e_k\mathbf{x}_k \quad (60)$$

where $\chi \in (0.95,1)$.

The value of $\mathrm{E}\left[\xi_k^4\right]$ is listed in Table 2.

Table 2. The value of $\mathrm{E}\left[\xi_k^4\right]$ for different distributes

| | Gaussian[15] | Uniform[15] | Binary[15] | Rayleigh[16] | Exponential[17] |
|---|---|---|---|---|---|
| $\mathrm{E}\left[\xi_k^4\right]$ | $3\sigma_\xi^4$ | $9\sigma_\xi^4/5$ | $\sigma_\xi^4$ | $8\sigma_\xi^4$ | $24\sigma_\xi^4$ |

Finally, the NCLMAT algorithm consists of Eq.(9)-(11), Eq.(54) and Eq.(59)-(60).

## 4. Simulation results

This section presents simulations in the context of system identification with various noises when the system is stationary or non-stationary to illustrate the accuracy of the NCLMAT algorithm. The length of the unknown coefficient vector $\mathbf{w}_k$ is $L$ ; the unknown coefficient is $\mathbf{w}_{opt}$ . The uncorrelated input signal $x_k$ is a Gaussian white noise signal with variance $\sigma_x^2 = 1$ and the correlated input signal $y_k$ is calculated by $y_k = 0.8y_{k-1} + x_k$ . In all the experiments, the coefficient vectors are initialized as a zero vector. $\mathrm{MSD}_k = 10\log_{10}\left(\left\|\mathbf{w}_k - \mathbf{w}_{opt}\right\|_2^2\right)$ is used to measure the performance of algorithms.

Fig. 1 compares the analytical (from Eq. (57)) and experimental results of the MSD for the proposed NCLMAT algorithm ($\beta$=0.001and $\gamma$=1000) for $x_k$ input with Uniform noise in stationary environment with an SNR=20 dB, close agreement between theory and simulation is obtained as depicted in Fig. 1. The unknown coefficient is $\mathbf{w}_{opt,k}$=[0.0227, 0.46, 0.688, 0.46, 0.227]$^T$. The results are obtained by Monte Carlo simulations with 10 independent running. Iteration number is 5000.

To understand clearly the improvement brought about the NCLMAT algorithm ($\beta$=0.001and $\gamma$=1000) when the $x_k$ input, the time-varying step-size $\alpha_k$ is depicted in Fig.2 for $x_k$ input with Uniform noise in stationary environment with an SNR=20 dB. The unknown coefficient is $\mathbf{w}_{opt,k}$ =[0.0227, 0.46, 0.688, 0.46, 0.227]$^T$. The time-varying step-size of the NCLMAT algorithm is shown in Fig.2. Similarly, the time variation of $\lambda_k$ of the NCLMAT algorithm is depicted in Fig. 3. As observed from this figure, $\lambda_k$ rises sharply and then decreases swiftly to reach zero. The results are obtained by Monte Carlo simulations with 10 independent running. Iteration number is 5000.

Fig. 4 displays the learning curves of the LMAT ($\mu$=0.01), NCLMAT ($\beta$=0.001and $\gamma$=10000) and ZNCLMAT ($\beta$=0.001and $\gamma$=1000) algorithms when the noise statistic is Exponential distributed signal with 2 for an SNR=10dB with $y_k$ input in nonstationary environment. The results are obtained by Monte Carlo simulations with 30 independent running. Iteration number is 5000. The unknown coefficient is $\mathbf{w}_{opt,k} = \mathbf{w}_{opt,k-1} + \boldsymbol{\upsilon}_k$ where $\mathbf{w}_{opt,k}$ =[0.0227, 0.46, 0.688, 0.46, 0.227]$^T$, $\boldsymbol{\upsilon}_k$ is an i.i.d. Gaussian sequence with $m_v = 0$ and $\sigma_v^2 = 0.01$. Here too, the excellent performance of the proposed NCLMAT algorithm is maintained and therefore a consistency in performance is achieved by the proposed NCLMAT algorithm.

Next, the choice of $\beta$ and $\gamma$ can also affect the performance of the NCLMAT

algorithm. Results in Fig. 5 for different values of $\beta$ and $\gamma$ show clearly that even in extreme cases. The system noise is Uniform distributed over the interval (-3, 3). The unknown coefficient is $\mathbf{w}_{opt,k}$ =[0.0227, 0.46, 0.688, 0.46, 0.227]$^T$ with $x_k$ input. The results are obtained by Monte Carlo simulations with 10 independent running. Iteration number is 10000. The MSD curves of the NCLMAT algorithm ($\beta$=0.001and $\gamma$=5000) with SNR=20dB. The performance of the NCLMAT algorithm did not deteriorate much. The best performance was achieved by increasing $\beta$ to 0.01 and decreasing $\gamma$ to 1000 for a constant steady-state misadjustment obtained by the NCLMAT algorithm.

Fig.6 summarizes the performance of the proposed NCLMAT algorithm ($\beta$=0.001 and $\gamma$=1000) in the fourth different noise environments with an SNR of 20dB with $y_k$ input in nonstationary environment. The results are obtained by Monte Carlo simulations with 30 independent running. Iteration number is 5000. The unknown coefficient is $\mathbf{w}_{opt,k} = \mathbf{w}_{opt,k-1} + \boldsymbol{\upsilon}_k$ where $\mathbf{w}_{opt,k}$ =[0.0227, 0.46, 0.688, 0.46, 0.227]$^T$, $\boldsymbol{\upsilon}_k$ is an i.i.d. Gaussian sequence with $m_v = 0$ and $\sigma_v^2 = 0.01$. From this figure that the best performance is obtained when the noise statistic is Rayleigh while the worst performance is obtained when the noise statistic is Uniform.

Moreover, to further test its performance, Fig.7 display the learning curves of the NCLMF ($\alpha$=0.001, $\beta$=0.0001 and $\gamma$=500) and NCLMAT algorithms ($\beta$=0.001and $\gamma$=10000) with SNR= 20dB when the noise statistic is Gaussian white noise with $m_\xi = 0$ and $\sigma_\xi^2 = 1$. The unknown coefficient is $\mathbf{w}_{opt,k}$ =[0.0227, 0.46, 0.688, 0.46, 0.227]$^T$. The correlated input signal was given by $y_k = 0.8y_{k-1} + x_k$.

Fig.8 display the learning curves of the NCLMAT algorithms ($\beta$=0.001and $\gamma$=20000) and ZNCLMAT ($\beta$=0.0003 and $\gamma$=2000) algorithms when the noise statistics are Exponential distributed signal with 2. The unknown coefficient is $\mathbf{w}_{opt,k} = \mathbf{w}_{opt,k-1} + \boldsymbol{\upsilon}_k$ where $\mathbf{w}_{opt,k}$ =[0.0227, 0.46, 0.688, 0.46, 0.227]$^T$, $\boldsymbol{\upsilon}_k$ is an i.i.d. Gaussian sequence with $m_v = 0$ and $\sigma_v^2 = 0.01$. The correlated input signal was given by $y_k = 0.8y_{k-1} + x_k$. The results are obtained by Monte Carlo simulations with 10 independent running. Iteration number is 5000. The unknown coefficient is $\mathbf{w}_{opt,k}$ =[0.0227, 0.46, 0.688, 0.46, 0.227]$^T$. The correlated input signal was given by $y_k = 0.8y_{k-1} + x_k$. Here too, the excellent performance of the proposed NCLMAT ($\beta$=0.001 and $\gamma$=2000) algorithm is maintained and therefore a consistency in performance is achieved by the proposed NCLMAT algorithm. Moreover, to further test its performance, the noise statistics are changed to Rayleigh distributed signal with 2 for an SNR=20 dB, and as depicted from Fig. 9, a similar behavior as the former ones is obtained by the proposed NCLMAT algorithm in this environment. The results are obtained by Monte Carlo simulations with 10 independent running. Iteration number is 5000. The tuning parameters for this part of the ZNCLMAT algorithm are $\beta$=0.003 and $\gamma$=100.

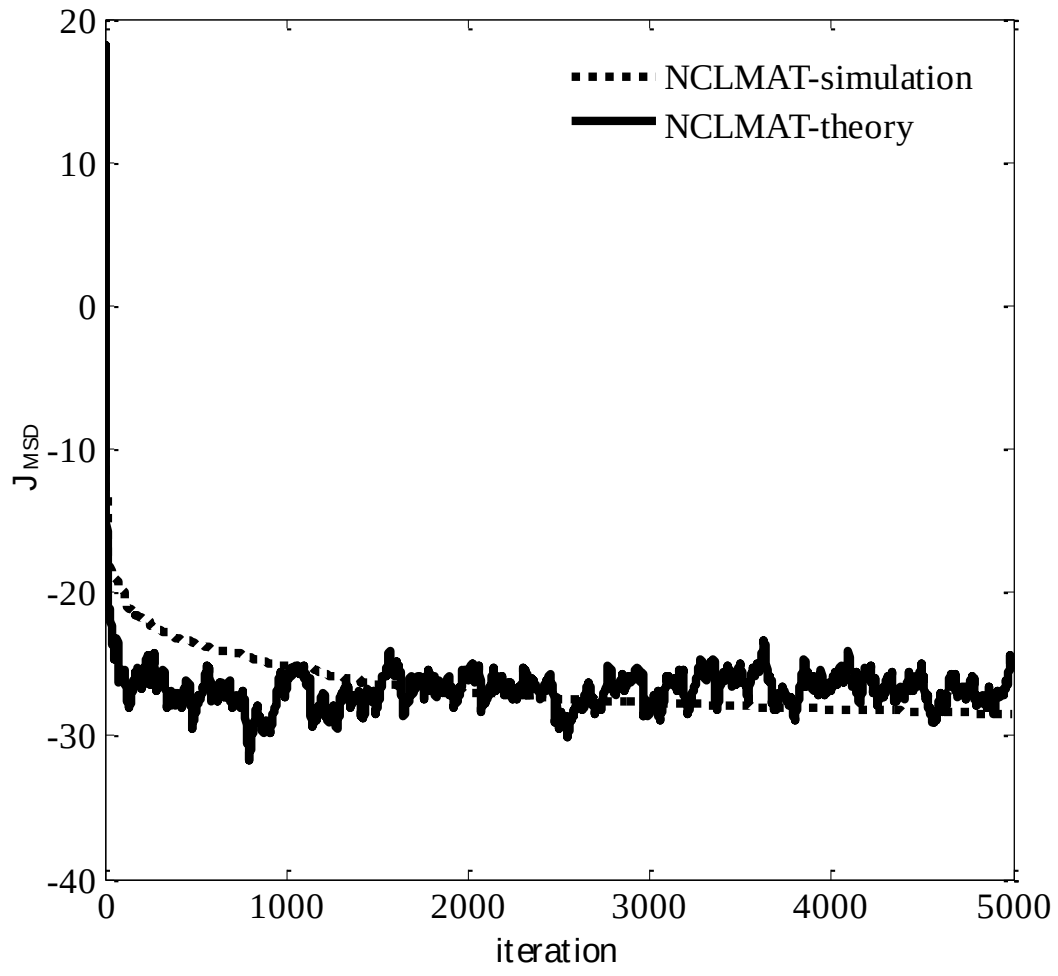

Fig.1. MSD learning curve of the NCLMAT algorithm

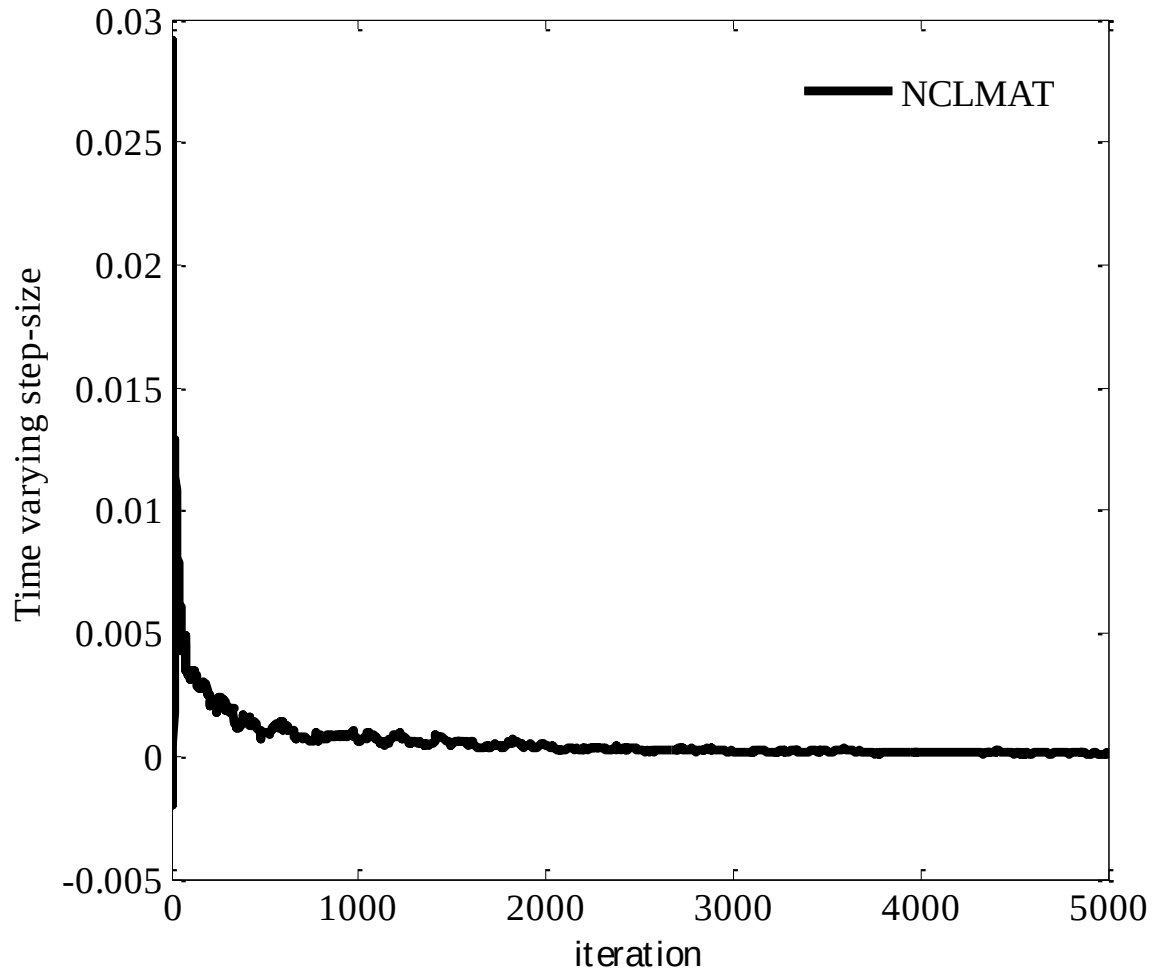

Fig.2. Behavior of time-varying step-size of NCLMAT algorithm

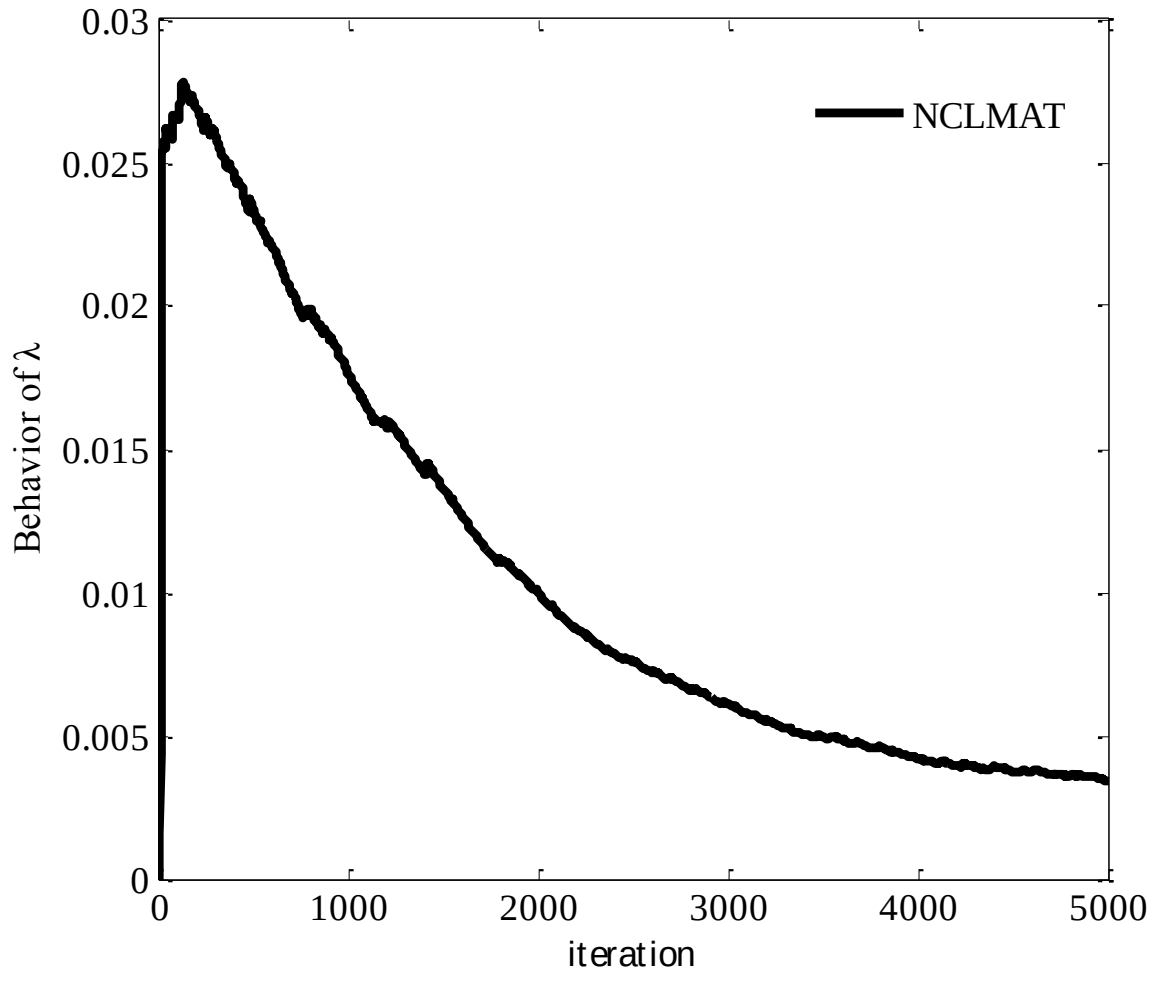

Fig.3. Behavior of $\lambda$ of NCLMAT algorithm

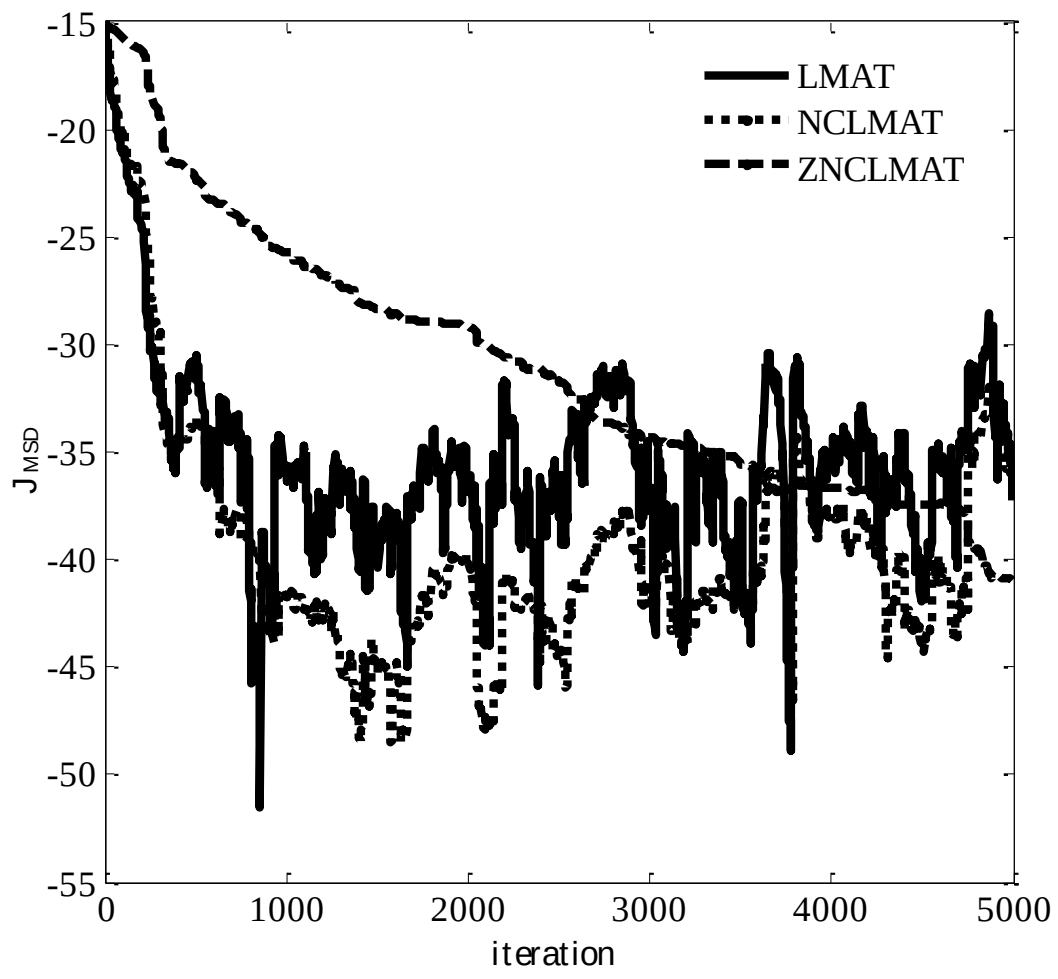

Fig.4. Learning curves for the LMAT, NCLMAT and ZNCLMAT algorithms

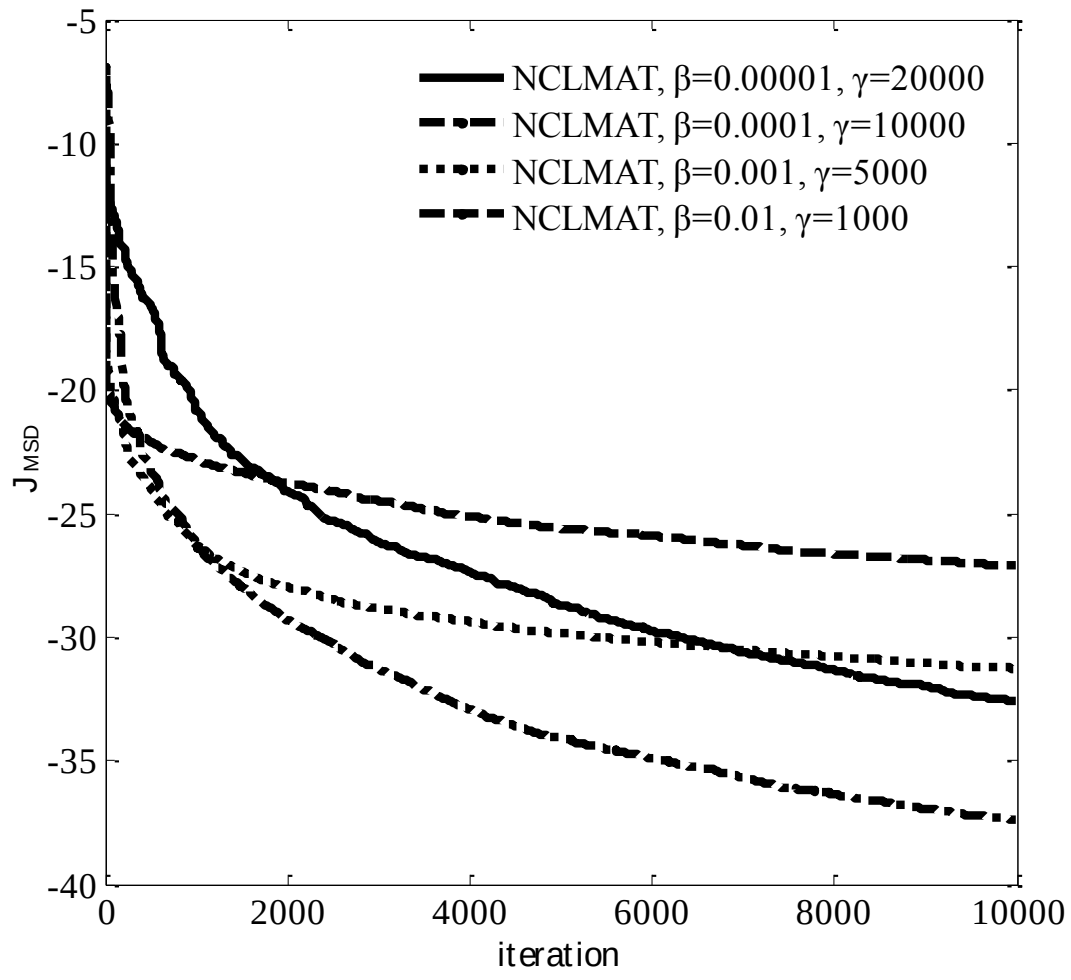

Fig.5. The effect of $\beta$ and $\gamma$ on the convergence behavior of the NCLMAT algorithm

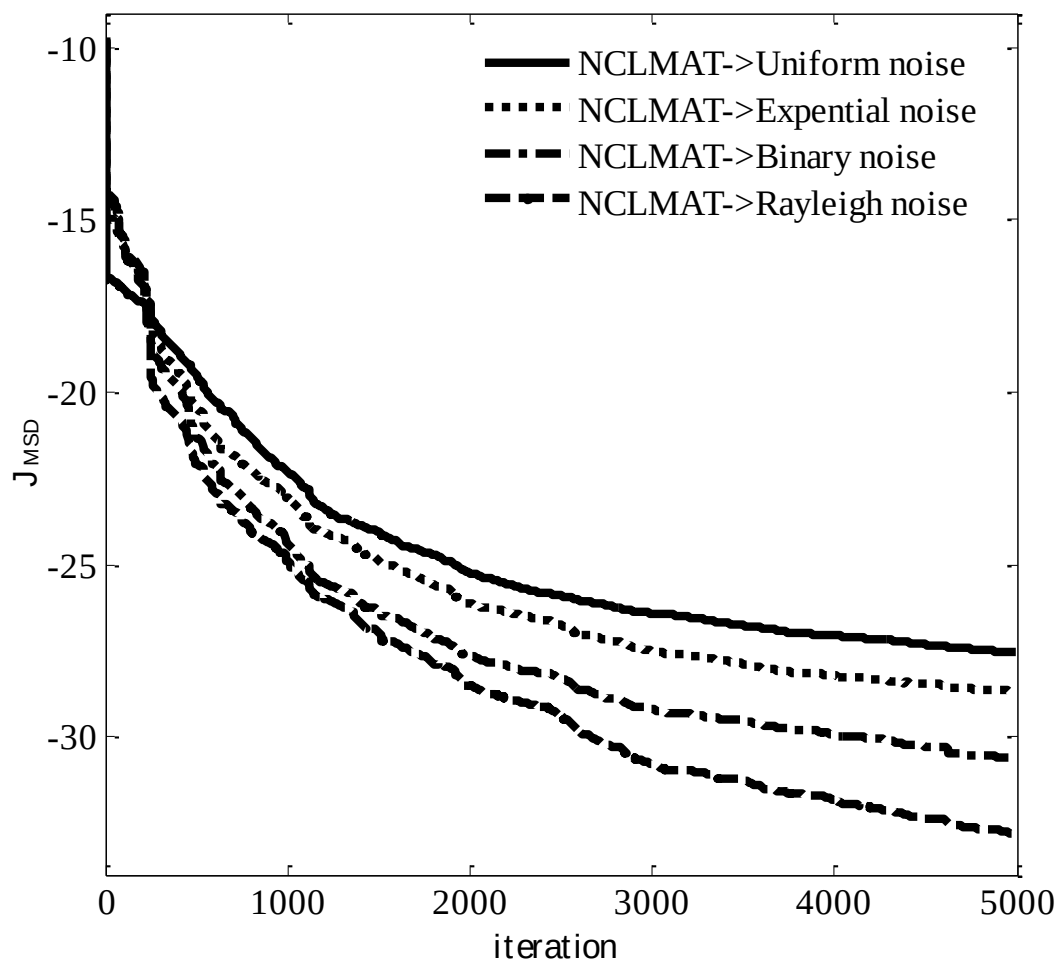

Fig.6. Learning curves for the NCLMAT algorithm in the fourth different noise environments

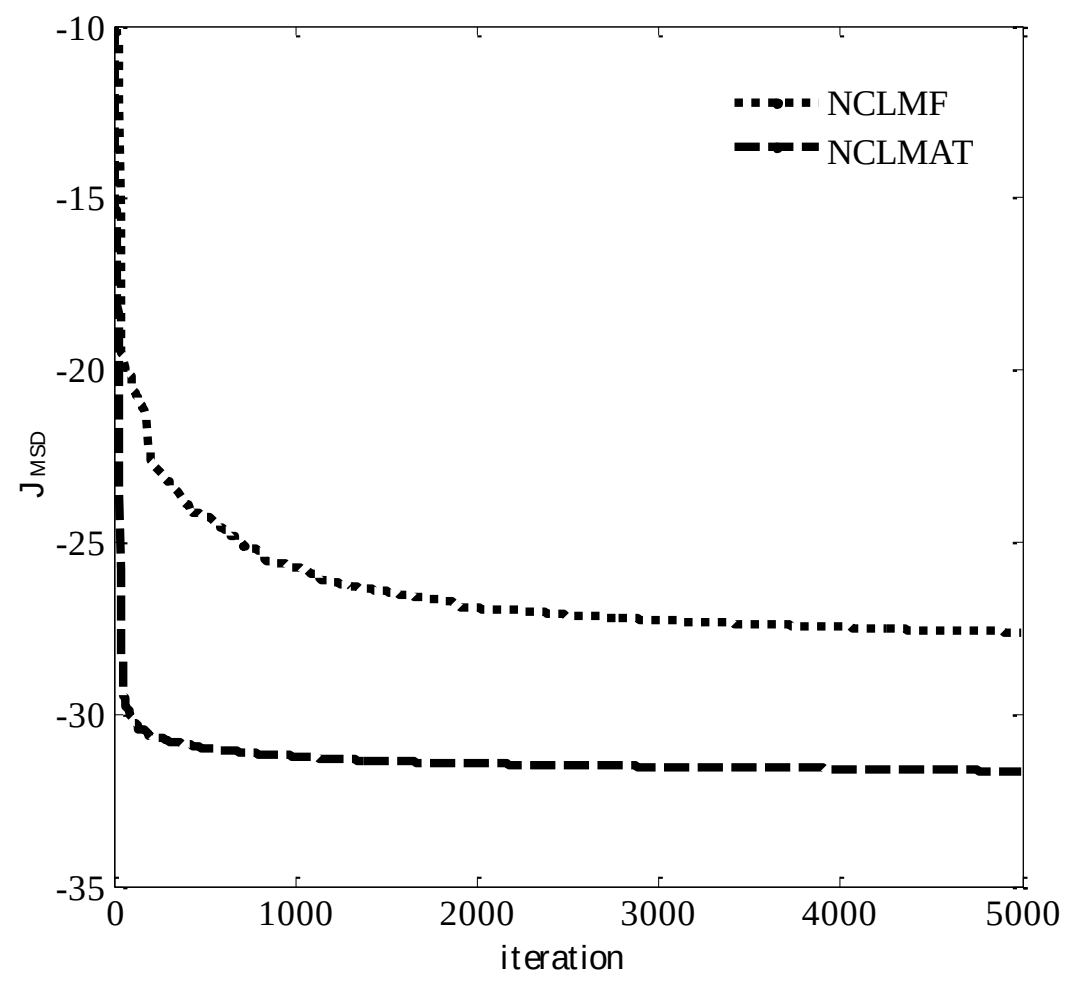

Fig.7. MSD learning curves of the NCLMF and NCLMAT algorithms.

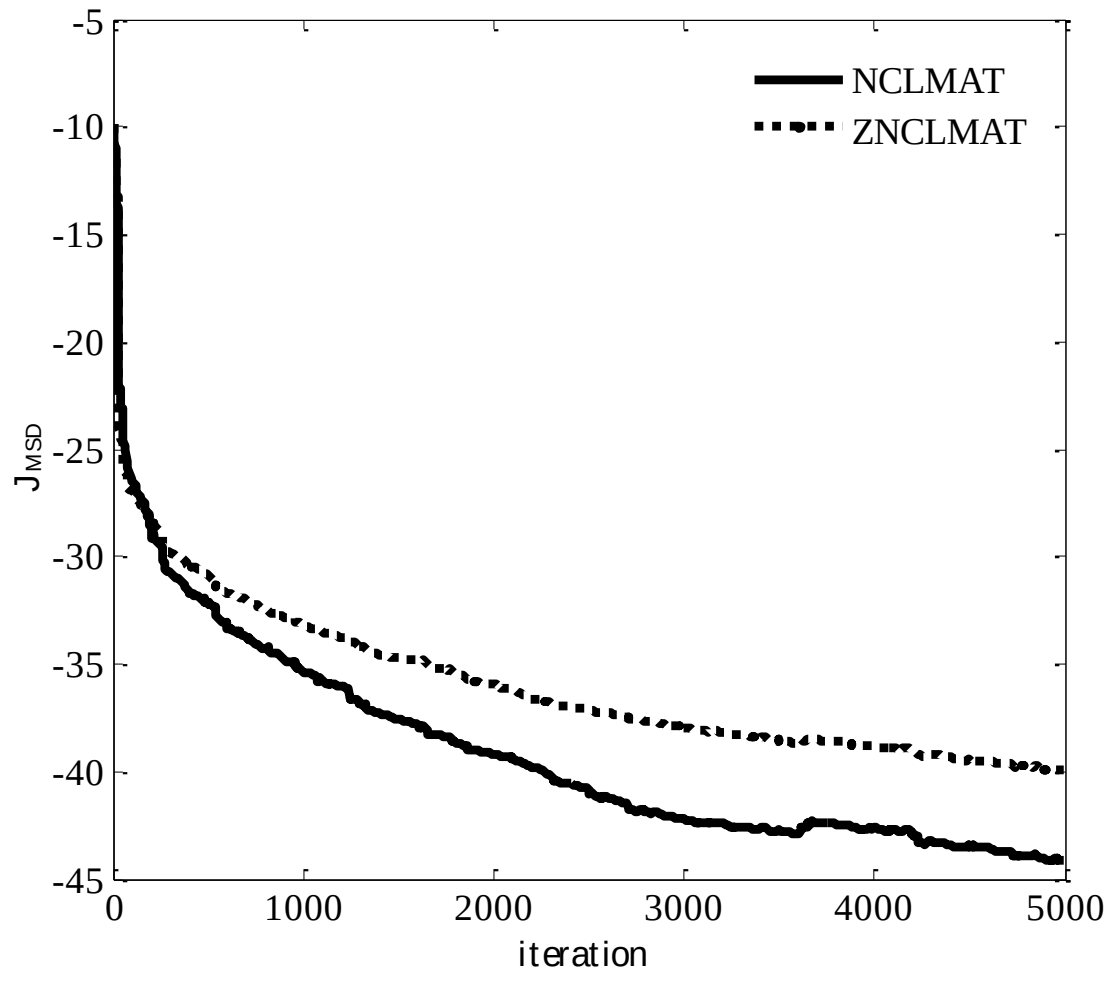

Fig.8. Behavior of time-vary step-size of NCLMAT and ZNCLMAT algorithms.

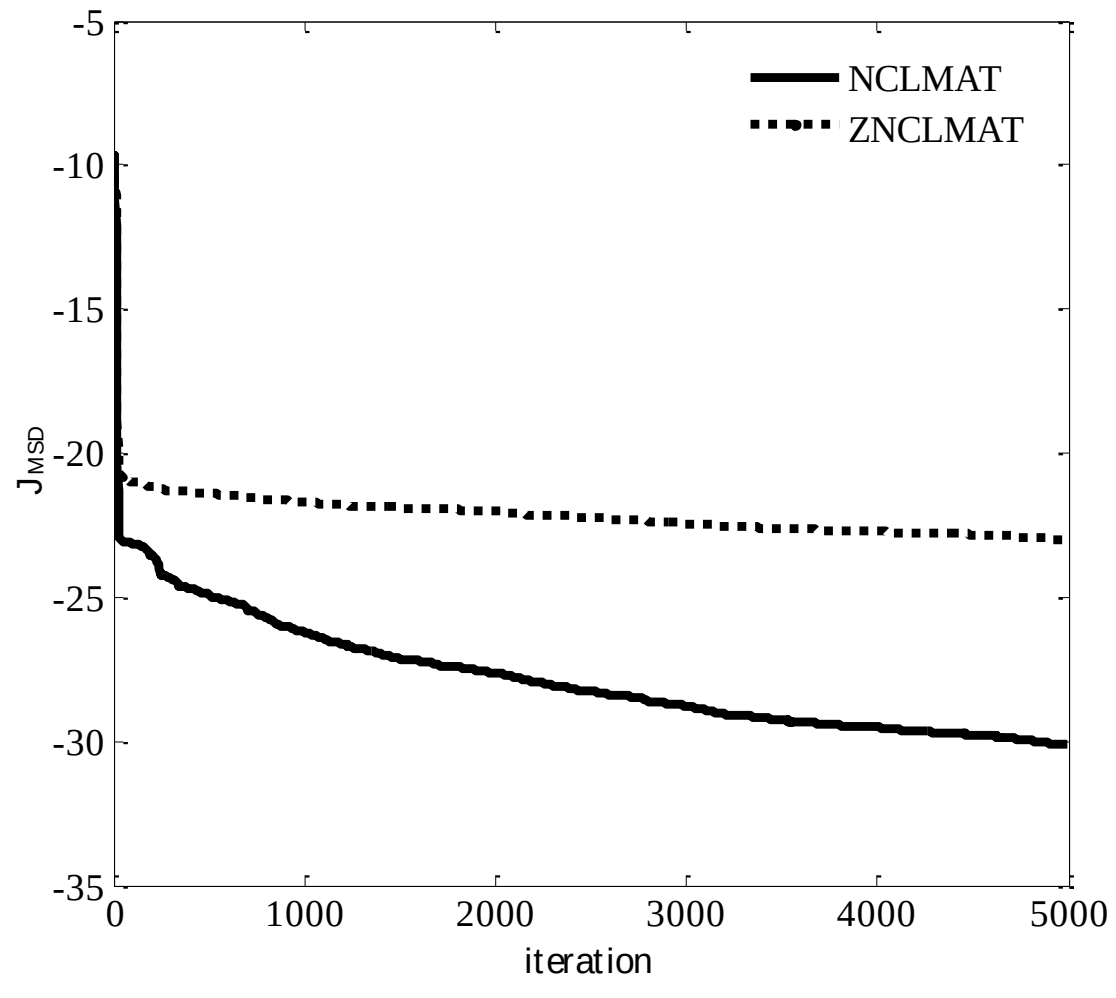

Fig.9. MSD learning curves of the NCLMF and NCLMAT algorithms.

## 5. Conclusion

In this paper, we proposed a noise constrained LMAT-type algorithm (NCLMAT) for FIR channel estimation and studied its performance both analytically and simulations. The NCLMAT algorithm is obtained by constraining the cost function of the standard LMAT algorithm to the third-order moment of the additive noise. The main aim of this work is to derive the NCLMAT algorithm, analyze optimized $\alpha$ and assess the performance in steady-state in different noise environments when the system is stationary or non-stationary. Compare with the NCLMF and LMAT algorithms, the NCLMAT enjoys a superior performance. The techniques introduced here may be applicable to plant identification.

**Acknowledgments** This work was supported by National Natural Science Foundation of China (61074120, 61673310) and the State Key Laboratory of Intelligent Control and Decision of Complex Systems of Beijing Institute of Technology.

## References

[1] H. Sayed. Adaptive Filters [M]. Hoboken, NJ, USA: John Wiley & Sons, Inc., 2008.
[2] P.S.R. Diniz, Adaptive Filtering (Fourth Edi.) [M]. Springer US, 2013
[3] Wei Y, Gelfand S B, Krogmeier J V. Noise-constrained least mean squares algorithm [J]. IEEE Trans. Signal Process. 49(9) (2001) 1961-1970.
[4] R. Kwong and E. W. Johnston. A variable step size LMS algorithm [J]. IEEE Trans. Signal Process. 40(7) (1992)1633-1642.
[5] Zhao H, Yu Y, Gao S, Zeng X, He Z. A new normalized LMAT algorithm and its performance analysis [J]. Signal Process. 105(12) (2014) 399-409.
[6] Zerguine A, Moinuddin M, Imam S A A. A noise constrained least mean fourth (NCLMF) adaptive algorithm [J]. Signal Process. 91(1) (2011) 136-149.
[7] Eweda E. Dependence of the Stability of the Least-mean Fourth Algorithm on Target Weights Non-Stationarity [J]. IEEE Trans. Signal Process. 62(7) (2014) 1634-1643.
[8] Lee Y H, Jin D M, Sang D K, Cho S H. Performance of least-mean absolute third (LMAT) adaptive algorithm in various noise environments [J]. Electron. Lett. 34(3) (1998) 241- 243.
[9] Guan S, Li Z. Nonparametric variable step-size LMAT algorithm [J]. Circ. Syst. & Signal Process., 2016:1-18.
[10] Cho S, Sang D K. Adaptive filters based on the high order error statistics[C]// Circuits and Systems, IEEE Asia Pacific Conference on. 1996 (1996)109-112.
[11] Hubscher P I, Bermudez J C M, Nascimento V H. A Mean-Square Stability Analysis of the Least Mean Fourth Adaptive Algorithm [J]. IEEE Trans. Signal Process. 55(8) (2007) 4018-4028.
[12] Cho S H, Mathews V J. Tracking analysis of the sign algorithm in nonstationary environments [J]. IEEE Trans. Acoustics Speech & Signal Processing, 38(12) (1990) 2046-2057.

[13] H.-C. Huang, J. Lee. A new variable step-size NLMS algorithm and its performance analysis. IEEE Trans. Signal Process. 60(4) (2012) 2055-2060.
[14] Han-Sol Lee, Seong-Eun Kim, Jae-Woo Lee, and Woo-Jin Song. A Variable Step-Size Diffusion LMS Algorithm for Distributed Estimation [J]. IEEE Trans. Signal Process. 63(7) (2015)1808-1820.
[15] Eweda E, Bershad N J. Stochastic Analysis of a Stable Normalized Least Mean Fourth Algorithm for Adaptive Noise Canceling With a White Gaussian Reference [J]. IEEE Trans. Signal Process. 60(12) (2012) 6235-6244.
[16] K. Hirano, Rayleigh Distribution (Wiley, London, 2014)
[17] Spiegel M R. Mathematical handbook of formulas and tables [M]. McGraw-Hill, 2012.